\documentclass{iopart}
\usepackage{graphicx}

\begin{document}

\newcommand{\remark}[1]{{\large\bf #1}}
   
\title{Negative-weight percolation}
\author{O Melchert and A K Hartmann}
\address{Institut f\"ur Physik, 
Universit\"at Oldenburg, 
Carl-von-Ossietzky Strasse, 
26111 Oldenburg, Germany}
\ead{melchert@theorie.physik.uni-oldenburg.de}

\begin{abstract}
We describe a percolation problem on lattices (graphs, networks), 
with edge weights 
drawn from disorder distributions that allow for weights (or distances) 
of either sign, i.e.\ including negative weights. We are interested whether
there are spanning paths or loops of total negative weight.
This kind of percolation problem 
is fundamentally different from conventional percolation problems,
e.g.\ it does not exhibit transitivity, hence no simple definition
of clusters, and several spanning paths/loops might coexist in the percolation
regime at the same time. Furthermore, to study this percolation problem
numerically, one has to perform a non-trivial transformation of the
original graph and apply sophisticated matching algorithms.
 
Using this approach,  we study the corresponding percolation
transitions on large square, hexagonal and cubic  
lattices for two types of disorder
distributions and determine the
critical exponents. The results show  that negative-weight percolation
is in a different universality class compared to conventional
bond/site percolation. On the other hand, negative-weight percolation
 seems to be related to the
ferromagnet/spin-glass transition of random-bond Ising systems, at
least in two dimensions.
\end{abstract} 


\section{Introduction \label{sect:introduction}}
\begin{figure}[t]
\centerline{
\includegraphics[width=0.6\linewidth]{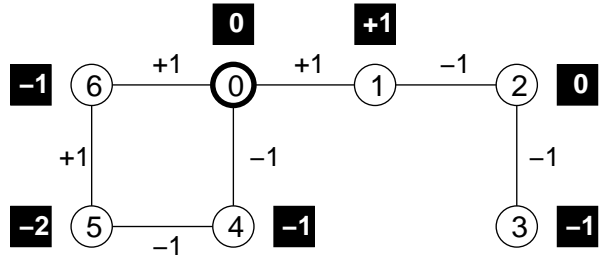}}
\caption{An example graph, exhibiting 7 nodes 0,$\ldots$,6
 with positive and negative edge weights (numbers
at lines). Numbers in black boxes denote the total weight 
of the minimum-weighted path from node 0 to the corresponding node.
\label{fig:example}}
\end{figure}

Percolation \cite{stauffer1994} is one of the most-fundamental problems
in statistical mechanics. Many phase transitions in physical systems
can be explained in terms of a percolation transition. The pivotal 
property of percolation is connectivity. One can describe this in terms
of weighted graphs (often also called networks)
with nonzero or zero weights, corresponding to occupied/connected or  
unoccupied/disconnected edges. 
Here, we extend the problem to negative weights,
in particular we study bond percolation. As an example, one can imagine
an agent traveling on a graph, who has, when traversing an edge,
 either to pay some resource 
(positive weight) or he is able (once) to  harvest 
some resource (negative weight). 
Paths including negative edge weights also appear in the
context of domain walls in random-bond Ising systems \cite{melchert2007}.
One percolation 
problem is whether there exists a path (or loop) spanning the full system
 with negative
total weight, such that each edge is traversed at most once.
Hence, the percolating objects are paths and it is sufficient
to look for minimum-weighted (or ``shortest'') paths (MWPs).
Percolation properties
of  string-like objects have been studied occasionally
\cite{engels1996,schakel2001,bittner2005,wenzel2005,antunes1998,pfeiffer2003},
but to our knowledge never allowing for negative weights.
One realizes immediately that the negative weights lead to properties
which are fundamentally different from conventional percolation. E.g.,
negative-weight percolation (NWP) lacks transitivity: If there is a valid
(negative-weight) path $S\to A\to B$ from $S$ to $B$ via $A$, it is possible
that the path $S\to A$ is not valid, as for the paths $0-1-2-3$/$0-1$ in figure
\ref{fig:example}. Hence, there is no simple definition of
percolation clusters in this case. Note that, since  strings
are thus considered, several percolating strings or loops might
coexist in the same sample. This we observe indeed, see below.

Here, we study NWP numerically. First we 
introduce the model. Then we outline the algorithm to obtain MWPs, 
since no standard shortest-path algorithm can be used. The results
for different types of two- ($2d$) and three-dimensional ($3d$) 
systems, in particular the
critical exponents describing the percolation transitions,
 show that NWP is fundamentally different compared to conventional 
percolation and from previous models with string-like percolation.
Thus, the study of percolation models exhibiting negative weights 
might lead to many new insights concerning the behavior of
disordered systems.

\section{Model and Algorithm\label{sect:model}}
\begin{figure}[t]
\centerline{
\includegraphics[width=0.3\linewidth]{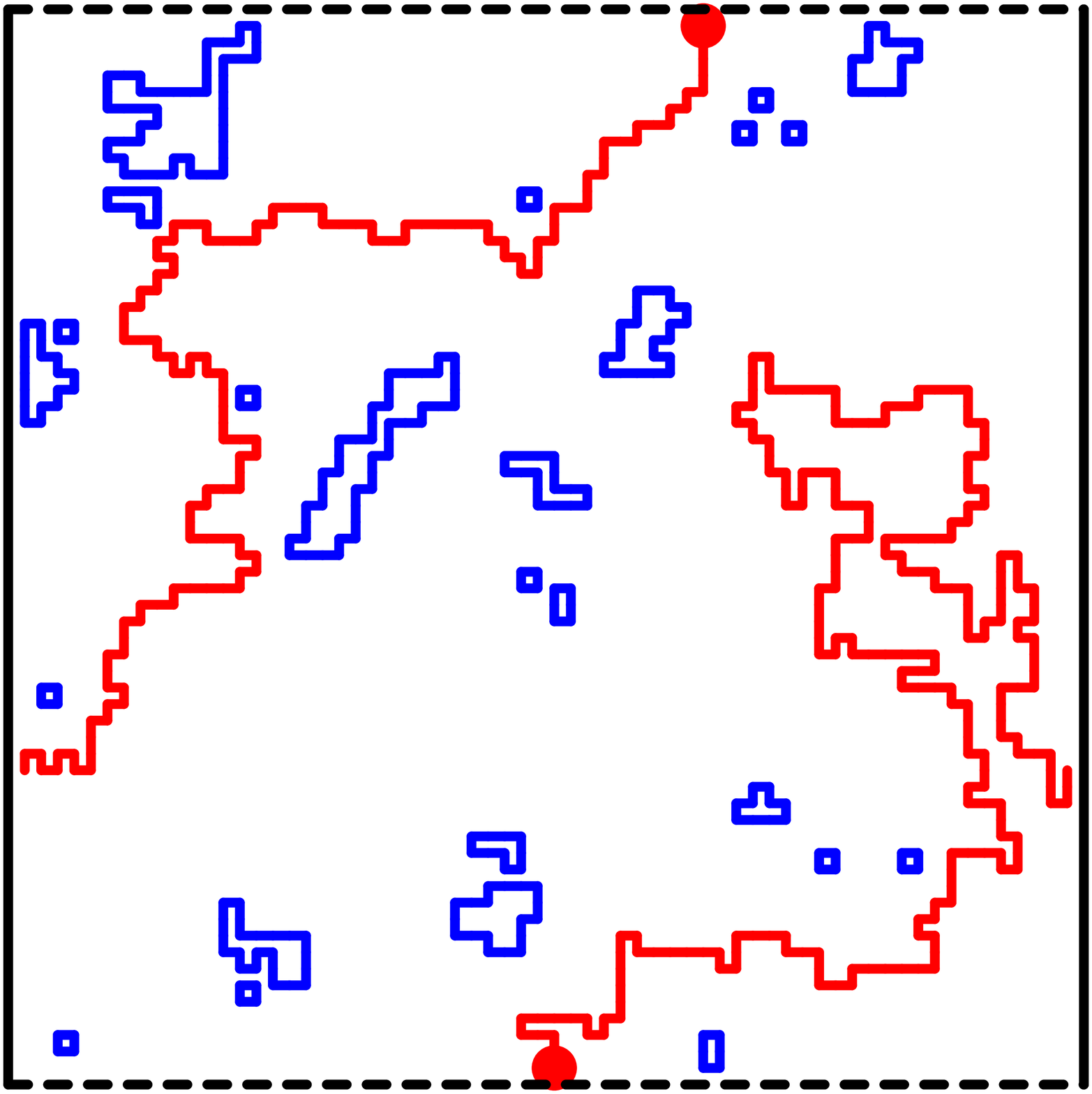}
\includegraphics[width=0.3\linewidth]{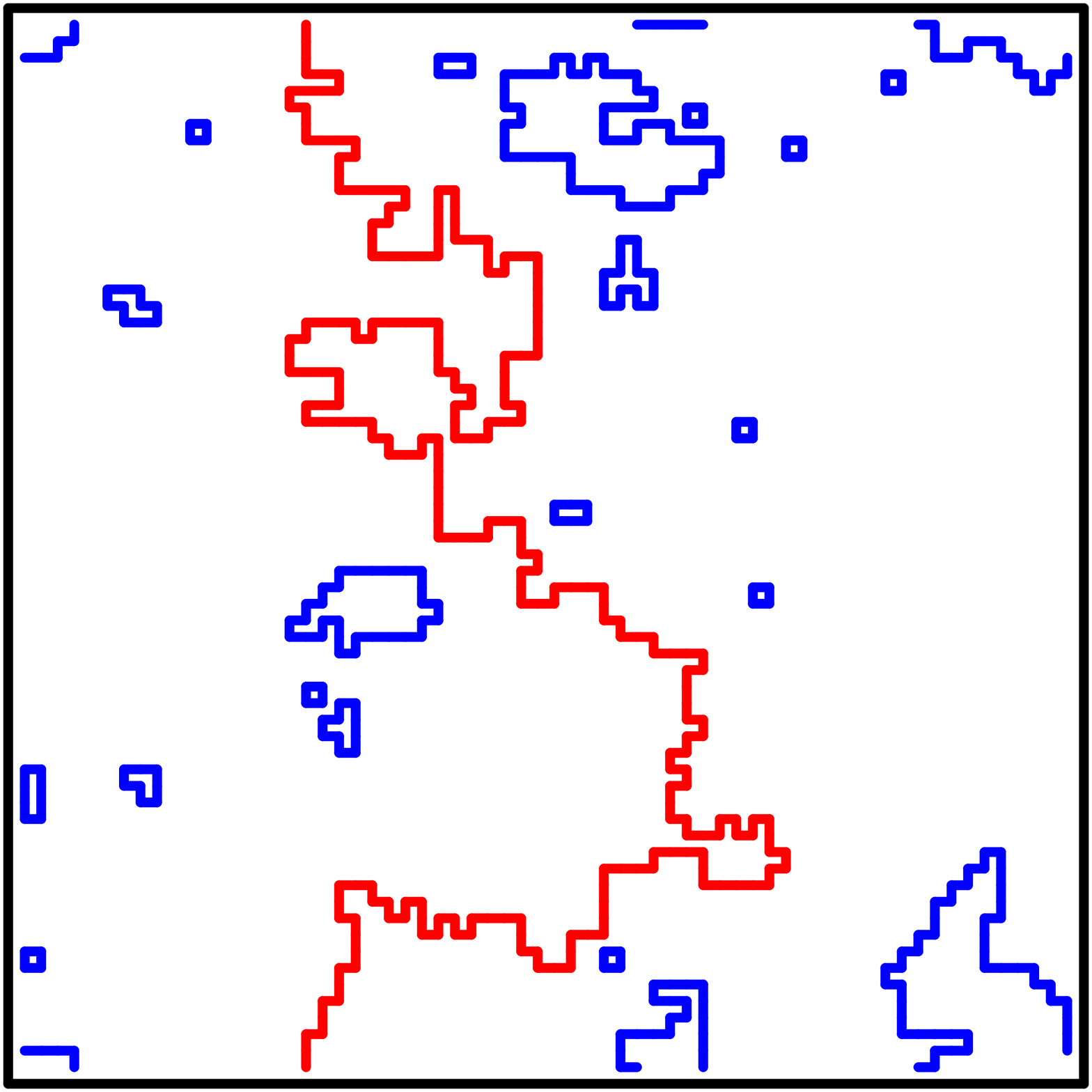} 
\includegraphics[width=0.3\linewidth]{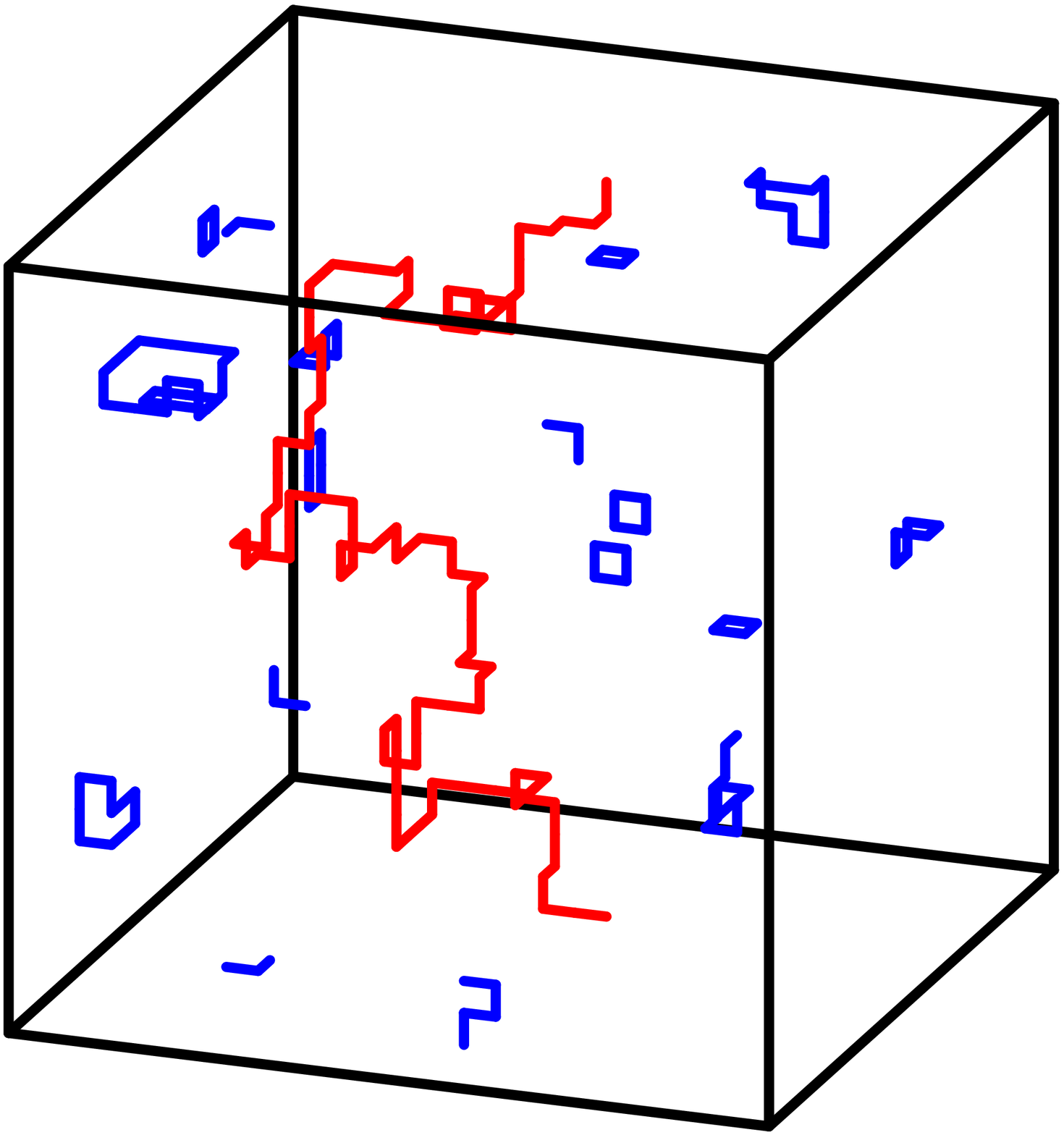}} 
\caption{
Left: MWP (red) in the presence of negative-weight 
loops (blue) for $L\!=\!64$, 
middle (right): $2d$ ($3d$) loop configuration for $L\!=\!64~(24)$. 
Spanning loops are colored red.
Each sample is taken slightly above the respective critical 
point $\rho_c$ for $\pm$J disorder. 
Dashed lines denote free BCs, solid lines indicate periodic BCs. 
\label{fig:samples}}
\end{figure}

We consider $2d$ and $3d$ 
lattice graphs $G=(V,E)$ with side length $L$, 
where adjacent sites $i,j\in V$ are joined by undirected edges $e=(ij)\in E$.
Weights $\omega(e)$ are associated with the edges, representing
quenched random variables.  We consider  
either bimodal ($\pm$J) or ``Gaussian-like'' (Gl)
distribution of the  edge weights, where $\rho$ denotes the fraction
of negative or Gaussian-distributed edge weights, respectively,
 among edges with unit weight (fraction $1-\rho$). 
The bimodal weights are taken to be $\pm1$ and the Gaussian 
weights have zero mean and unit width.
These weight distributions explicitly allow for loops ${\mathcal L}$ 
with negative
weight, given by 
$\omega_{\mathcal L}\!\equiv\!\sum_{e\in {\mathcal L}}\omega(e)$. 
For any non-zero
value of $\rho$, a sufficiently large systems will exhibit at least small
loops with negative total weight.
  We investigate (i) MWPs in the
presence of negative weighted loops on square lattices with periodic
boundary conditions (BCs) in one and free BCs  in the other
direction. The path ends are allowed to terminate on the free
boundaries.  Further, we study (ii) minimum-weighted configurations of
negative-weighted loops on square, $2d$ hexagonal and cubic 
lattices with fully periodic BCs, see figure \ref{fig:samples}.
 In either case we find critical values of
$\rho$ above which  (i) negative-weight 
paths appear that span the lattice across the
direction with  periodic BCs or (ii) percolating loops emerge, that
span the lattice along some direction.


Since we are looking for MWPs on undirected graphs in the presence
of negative weights, traditional ``shortest path'' algorithms cannot
be applied. The reason is that for applying these traditional algorithms a
special condition must hold:
For the
distance $d(i)$ of any shortest path from a source $0$ to $i\neq 0$,
$d(i)=\min_{j\in N(i)} (d(j)+\omega(j,i))$ holds, where
$N(i)$ denotes the set of neighbors of $i$. This equation is not fulfilled
in our case, as can be seen from figure \ref{fig:example} for node $4$. It
has minimum distance $-1$ to node $0$, 
but it is connected to node 5 via an edge of
weight $\omega(5,4)\!=\!-1$ and $d(5)\!=\!-2$. 
Hence, a different approach has to be applied.
We determine MWPs and loop configurations through an 
appropriate transformation of the original graph, detailed in \cite{ahuja1993}, 
and obtaining a minimum-weighted perfect matching (MWPM) 
\cite{cook1999,opt-phys2001}, by using exact 
combinatorial optimization algorithms.

\begin{figure}[t]
\centerline{
\includegraphics[width=0.94\linewidth]{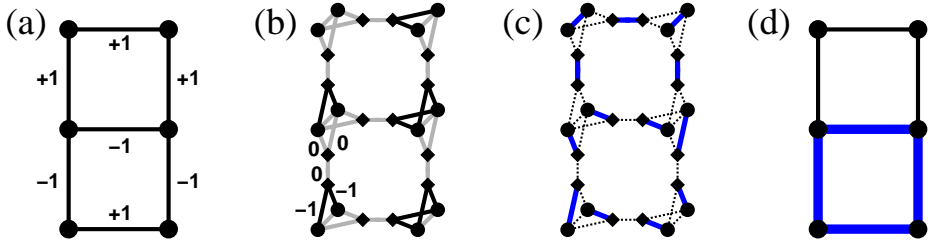}}
\caption{ Illustration of the mapping procedure.
(a) original lattice $G$ with edge weights, 
(b) auxiliary graph $G_{\rm A}$ with proper weight assignment: black 
edges carry the same weight as the respective edge in the original
graph and gray edges carry zero weight. 
(c) minimum-weight perfect matching: bold edges are matched 
and dashed edges are unmatched.
(d) loop configuration (bold edges) that corresponds to the MWPM.
\label{fig:alg}}
\end{figure}

Here we give a concise description of the mapping,
pictured as 3-step procedure illustrated in figure \ref{fig:alg}:

(1) each edge, joining 2 sites on $G$, is replaced by a path of 3 edges. 
Therefore, 2  ``additional'' sites have to be introduced for each
edge of $G$. The original 
sites are then ``duplicated'' along with their incident edges. 
For each of these pairs of  of duplicated sites, one additional 
edge is added that connect the two sites of a pair.
The resulting auxiliary graph $G_{{\rm A}}$ is depicted in figure 
\ref{fig:alg}(b), where additional sites appear as squares and duplicated 
sites as circles. Figure \ref{fig:alg}(b) also illustrates the weight 
assignment on $G_{{\rm A}}$.
A more extensive description can be found in \cite{melchert2007}.

(2) a MWPM on the auxiliary graph is determined \cite{comment_cookrohe}. 
A MWPM is a minimum-weighted subset $M$ of the edges contained in 
$G_{\rm A}$, such that each site of $G_{\rm A}$ is met by one edge in $M$. 
This is illustrated in figure \ref{fig:alg}(c), where the bold edges 
represent $M$ for the given weight assignment. The dashed edges are 
unmatched.
Due to construction, the auxiliary graph consists of an even number of sites 
and since there are no isolated sites, it is guaranteed that a perfect
matching exists.

(3) finally it is possible to find a relation between the matched edges $M$ 
on $G_{\rm A}$ and a configuration of negative-weighted loops 
$\mathcal{C}$ on $G$ by 
tracing the steps of the transformation (1) back. Note that each edge 
contained in $M$ that connects an additional site (square) to a duplicated 
site (circle) corresponds to an edge on $G$ that is part of a loop, see 
figure \ref{fig:alg}(d). More precise, there are always two such edges in $M$ 
that correspond to one edge on $G$. All the edges in $M$ that connect like 
sites (i.e.\ duplicated-duplicated, or additional-additional) 
carry zero weight and do not contribute to a loop on $G$.
Afterwards a depht-first search can be used to explore $\mathcal{C}$ and 
to determine the properties of the individual loops. For the weight
assignment in figure \ref{fig:alg}(a), there is only one loop with 
weight $\omega_{\mathcal L}\!=\!-2$ and length $\ell=4$.

Note that the result of the calculation is a collection $\mathcal{C}$ of loops
(and one path for (i), see below), such that the total loop
weight is minimized. Hence, one obtains a global collective optimum
of the system.
Clearly, all loops that contribute to $\mathcal{C}$ possess a negative weight. 
Note that $\mathcal{C}$ can be empty and that sub paths are 
neither allowed to intersect nor to terminate at some site 
within the lattice. 

%
%
So as to induce a MWP, as in problem (i), special care is needed. 
We have to allow the paths explicitly to terminate at a certain node. 
Therefore, two extra sites are introduced to the graph. 
One extra site is connected to each of the free boundaries by adding edges
 with weight zero to the transformed graph 
that join the site with the sites of the respecting 
boundary. Any subsequent MWPM will contain a path of minimal weight, 
joining the extra sites. This does not necessary coincide with the 
``shortest'' path, as explained above, but yields the 
minimal-weighted path in the presence of negative weighted loops. 
In contrast to the loops, MWPs can carry a positive weight in principle, 
but, as we will see, this will not be the case close to and beyond the
percolation transition we are interested in.


%
%
\begin{table}[b!]
\caption{\label{tab:crit_exp}
Critical points and exponents for the paths and loops.  From
left to right: Disorder type (P: path, L: loop,  $\pm$J: bimodal, 
Gl: Gaussian-like), 
lattice geometry (sq: square, hex: hexagonal, cu: cubic), 
critical point $\rho_c$, critical exponent of the correlation 
length $\nu$,  percolation strength $\beta$, exponent $\gamma$, 
Fisher exponent $\tau$ and fractal dimension $d_f$ at criticality.} 
\begin{indented}
\item[]\begin{tabular}[c]{@{}llllllll}
\br
Type        & geom.  	   & $\rho_c$ 	& $\nu$    & $\beta$  	& $\gamma$ & $\tau$  & $d_f$   \\
\mr
P$\pm$J & $2d$ sq & 0.1032(5)	& 1.43(6)  & 1.03(3)	& 0.76(5)  & 2.52(8)  	&  1.268(1) \\ 
L$\pm$J & $2d$ sq & 0.1028(3) 	& 1.49(9)  & 1.09(8) 	& 0.75(8)  & 2.58(6)  	& 1.260(2) \\ 
L$\pm$J & $2d$ hex& 0.1583(6) 	& 1.47(9)  & 1.07(9) 	& 0.76(8)  & 2.59(2)  	& 1.264(3) \\ 
L-Gl    & $2d$ sq & 0.340(1) 	& 1.49(7)  & 1.07(6) 	& 0.77(7)  & 2.59(3)  	& 1.266(2) \\ 
L$\pm$J & $3d$ cu & 0.0286(1) 	& 1.02(3)  & 1.80(8)	& --       & 3.5(3)	&  1.30(1) \\ 
\br
\end{tabular}
\end{indented}
\end{table}

\begin{figure}[t]
\centerline{ \includegraphics[width=0.98\linewidth]{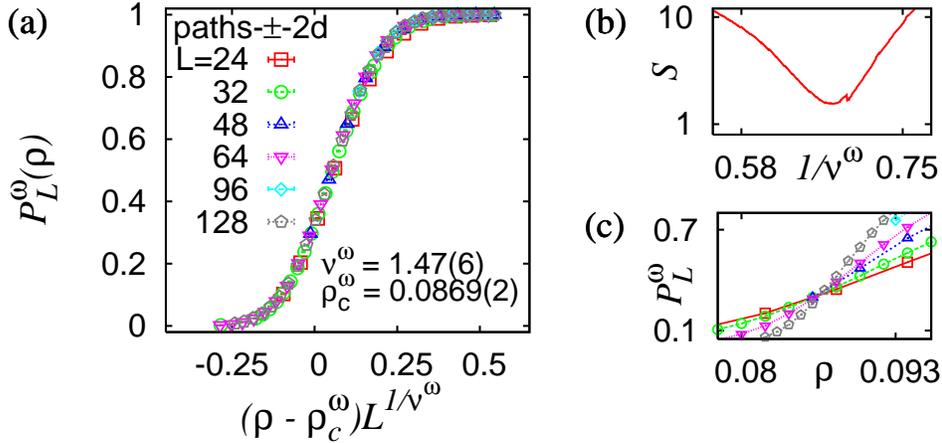} }
\caption{ Results for $2d$ $\pm$J paths: 
(a) Rescaled probability $P_{L}^{\omega}$ that 
the path weight is negative for different system sizes $L$,  
(b) illustrates the quality $S$ of the critical exponent 
$\nu^{\omega}$ and  
(c) shows the unscaled data near
$\rho_c^{\omega}$.
\label{fig:path_nu}}
\end{figure}  

\begin{figure*}[t]
\centerline{\includegraphics[width=0.7\linewidth]{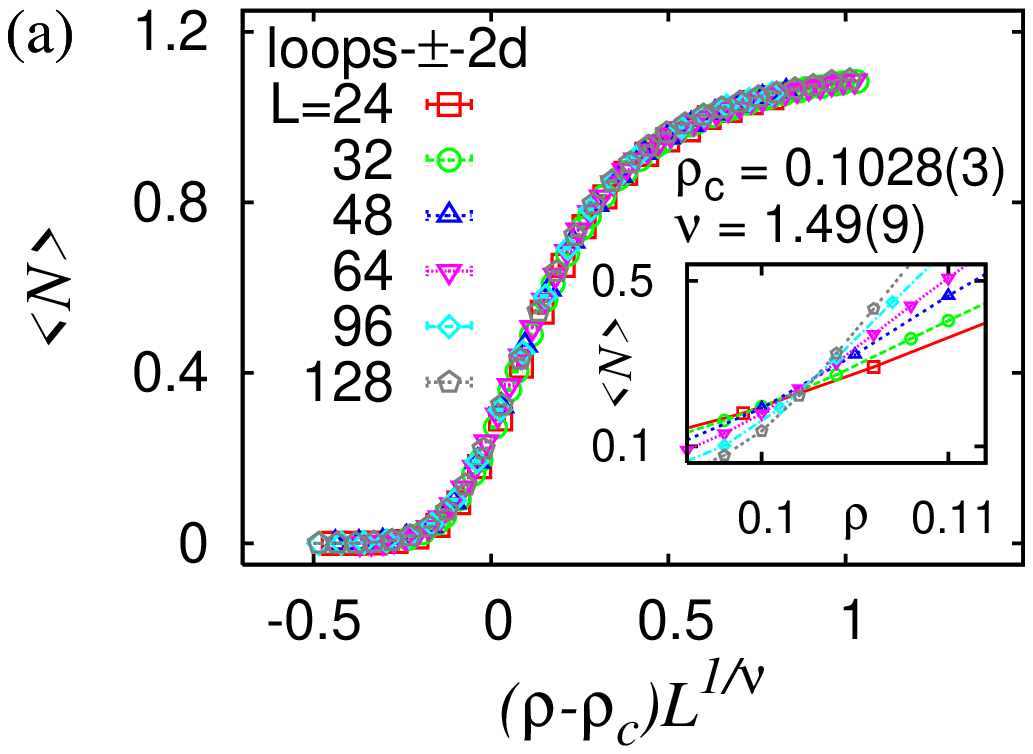}}
\centerline{\includegraphics[width=0.7\linewidth]{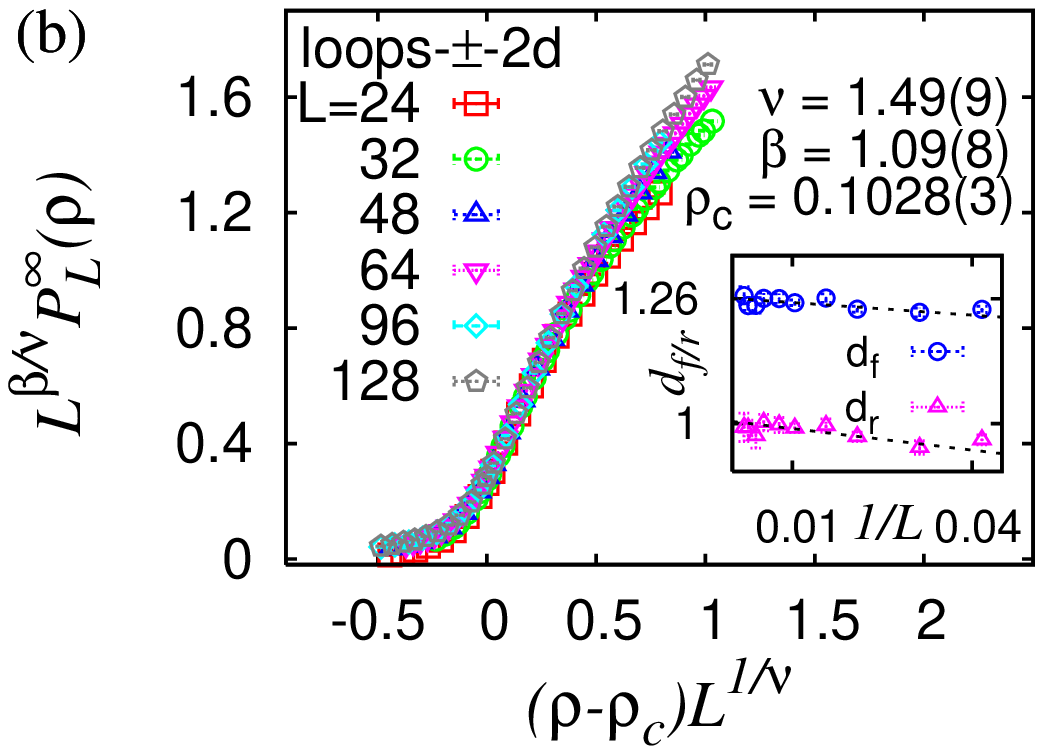}}
\caption{ Results for the $2d$ $\pm$J loops: (a) Mean
number $\langle N \rangle$ of percolating loops for different  system
sizes $L$. The inset shows the unscaled data near $\rho_c$, (b)
rescaled probability $P_{L}^{\infty}(\rho)$ that an edge belongs to a
percolating loop.  The inset shows the local slopes as function of
$1/L$, describing   the scaling of $\langle \ell \rangle$ at $\rho_c$,
\label{fig:loops_2d}}
\end{figure*}  

\section{Minimal-weighted paths \label{sect:mwp}}
Within this problem, the aforementioned collection $\mathcal{C}$
consists of a set of loops, which might be empty, and a MWP spanning
the lattice between the free boundaries. Since the  MWP is anyway of
$O(L)$ in the vertical direction by construction, we call the path
percolating, if its projection on the horizontal axis (i.e.\ its {\em
roughness} $r$)  covers the full systems. We studied $2d$ lattices
with $\pm$J disorder and sampled over up to
$3.5\!\times\!10^{4}~(L\!=\!128)$ realizations of the disorder to
perform averages, subsequently denoted by $\langle \ldots \rangle$.
Firstly, we investigate the percolation probability.  For a very small
fraction $\rho$ of negative edge weights, the path will cross  the
lattice in a rather direct fashion, since overhangs are likely to
increase the weight of the path and hence the weight of the whole
configuration $\mathcal{C}$.  Increasing $\rho$ also increases the
spanning probability $P_{L}^{\rm{x}}(\rho)$,  i.e.\ the probability of
the horizontal extension to be $O(L)$. 
It is expected to scale as
$P_{L}^{\rm{x}}(\rho)\!\sim\!f[(\rho-\rho_c) L^{1/\nu}]$, where the
critical exponent $\nu$ describes the divergence of the correlation
length at the critical point $\rho_c$ and $f$ is a universal function.
A data collapse, obtained using the method described in
\cite{s_value}, yields $\rho_c$ and $\nu$  as listed in table
\ref{tab:crit_exp} with a ``quality'' $S\!=\!0.79$ of the scaling
assumption.  We want to point out that the value of $\nu$ found here
is clearly distinct from the value $4/3$  of conventional bond
percolation, that would give a much worse quality of $S\!=\!5.20$ (see
also figure \ref{fig:path_nu}(b)).
Note that many negative weights are needed to allow the MWP to
percolate.  And indeed, the path weight becomes negative
above $\rho_c^{\omega}\!=\!0.0869(2)$ with a probability that
approaches unity quickly. Therein, the finite-size
behavior is described by an exponent $\nu^{\omega} $ consistent with
the value $\nu$ stated above, see figure \ref{fig:path_nu}.

Further note that $P_{L}^{\rm{x}}(\rho)\!<\!1$  even for large $\rho$,
where the spanning probability seems to saturate slightly above $0.8$.
The reason is, as already mentioned, that we actually do not optimize
the length of a  single path but the weight of the whole configuration
$\mathcal{C}$.
However, this behavior is clearly distinct from conventional
percolation theory.

Next, we consider the probability $P_L^{\infty}\!\equiv\!\langle \ell
\rangle/L^{d}$ that a bond belongs to  the path, where $\langle \ell
\rangle$ is the mean path length. It exhibits the finite-size scaling
behavior  $P_{L}^{\infty}(\rho)\!\sim\! L^{-\beta/\nu} f[(\rho-\rho_c)
L^{1/\nu}]$, where $\beta$ signifies the  percolation strength
\cite{stauffer1994}. Here, we fixed $\rho_c$ and $\nu$ as obtained
above and  determined the value of $\beta$ ($S\!=\!0.97$) listed in
table \ref{tab:crit_exp}. Adjusting all three parameters in 
the above scaling assumption yields $\rho_c\!=\!0.1027(1)$, 
$\nu\!=\!1.45(4)$ and $\beta\!=\!1.06(2)$ with a quality $S\!=\!0.76$. 
Note that the values agree within the errorbars.

We also determined some quantities (see also table \ref{tab:crit_exp}) 
just at $\rho_c=0.1032(5)$ with additional
simulations up to $L=512$ ($2\times 10^3$ samples).  We studied the
associated finite-size susceptibility  $\chi_L\!=\!L^{-d}(\langle
\ell^{2}\rangle\! -\!\langle \ell \rangle^2)$.  Its finite
size-scaling at $\rho_c$ can be described using the exponent $\gamma$
via $\chi_L\!\sim\!L^{\gamma/\nu}$.  Furthermore, the mean path length
shows the critical behavior $\langle \ell \rangle\!\sim\!L^{d_f}$,
where $d_f$ denotes the  fractal exponent of the paths.
Here, calculations at $\rho_c\!=\!0.1027(1)$ would result in a slightly 
smaller value of $d_f$.
%
We further find the roughness exponent $d_r$ from  $\langle r
\rangle\!\sim\!L^{d_r}$ to be compatible with unity.  Note that the
obtained exponents satisfy the scaling relations
$d_f\!=\!d\!-\!\beta/\nu$  and $\gamma\!+\!2\beta\!=\!d\nu$.
We also measured the mean path weight $\langle \omega_p \rangle$ at
the percolation point $\rho_c$  and found that $\langle \omega_p
\rangle \!\sim\!\langle \ell \rangle$ for  $L\!\rightarrow\!\infty$
seems to hold.

We can further probe $\mathcal{C}$ to not only investigate 
the MWP but to  yield
exponents that describe the small loops.  A detailed study of the
scaling behavior (length $\ell$ as function of the spanning length $R$,
not shown here) at $\rho_c$ shows that they seem within 
error bars to exhibit {\em the same} fractal dimension  
$\tilde{d}_f\!=\!d_f$ as the MWPs.  The loops also
exhibit a mean (negative) weight that increases linearly with the
loop length $\ell$.  
Further, one expects the distribution $n_\ell$ of loop lengths $\ell$
to exhibit an algebraic scaling $n_\ell\!\sim\!\ell^{-\tau}$, where $\tau$
signifies the Fisher exponent. 
Here, finite-size effects and the presence of the path lead to a 
suppression of large loops and thus to a deviation from the expected 
scaling behavior already for rather small values of $\ell$. 
In this question, we find a most reliable value $\tau$ if we account for
corrections to scaling via $n_\ell\!\sim\!\ell^{-\tau}/(1+b\ell^{\omega})$, 
see table \ref{tab:crit_exp} and inset of figure \ref{fig:len_distrib}.
In principle, at criticality, the Fisher exponent and 
fractal dimension are related through the scaling relation 
$\tau\!-\!1\!=\!d/d_f$. This would lead us to expect $\tau\!\approx\!2.58$.

Finally, in $2d$ we can associate a cluster with each loop. The
respective volume $v$ is measured as the number of enclosed
plaquettes. It scales as $\langle v\rangle\!\sim\!R^{\tilde{d}_v}$
with $\tilde{d}_v=2.00(1)$, revealing the compact nature of the loops
interior.
%
%

\section{Minimal-weighted loop configurations \label{sect:mwl}}
Here, we studied $2d$ (hex, $3d$) lattices with size up to $L\!=\!128$ ($192$, $48$)
and  $4.5\!\times\!10^{4}$ ($1.2\!\times\!10^{4}$, $1.6\!\times\!10^{4}$) samples.  First, we
analyze the largest loop for a given realization of the disorder.
We find the linear extensions of the loops  by projecting
it onto  all perpendicular axes. A loop is said to span the lattice,
if its projection completely covers at least one axis.
%
%
From the probability $P_{L}^{\rm{s}}(\rho)$ that a loop spans the
lattice  we estimate $\rho_c$ and $\nu$ listed in table
\ref{tab:crit_exp}. The qualities of the  scaling law were found to be
$S^{\pm}_{\rm{2d}}=1.09$, $S^{\pm}_{\rm{hex}}=0.79$, $S^{\rm{Gl}}_{\rm{2d}}=0.91$ and
$S^{\pm}_{\rm{3d}}=1.9$.  
%
Further, it is interesting to note that the mean number of spanning
loops $\langle N \rangle$  satisfies a similar scaling relation as the
percolation probability, see figure \ref{fig:loops_2d}(a),  governed  by
the same values for $\rho_c$ and $\nu$ as $P_{L}^{\rm{s}}(\rho)$.  In
either case we find $\langle N \rangle\!>\!1$ for large values of
$\rho$, as mentioned already in the introduction, see also \cite{pfeiffer2003}.
As above, the probability $P_{L}^{\infty}(\rho)$ that an edge belongs
to the percolating loop  can be used to determine the exponent
$\beta$, see figure \ref{fig:loops_2d}(b) and table \ref{tab:crit_exp}.
Herein, the qualities of the scaling law were
$S^{\pm}_{\rm{2d}}\!=\!1.88$, $S^{\pm}_{\rm{hex}}=0.32$, 
$S^{\rm{Gl}}_{\rm{2d}}\!=\!1.16$ and
$S^{\pm}_{\rm{3d}}\!=\!2.08$.
Interestingly, the values of $\rho_c$, $\nu$ and $\beta$ for the
$2d$ loops with  $\pm$J disorder are reasonable close to those of the
$2d$ random-bond Ising model with  $\rho_c\!=\!0.103(1)$,
$\nu\!=\!1.55(1)$ and $\beta\!=\!0.9(1)$, see  \cite{amoruso2004} and
references therein. This probably means that the $2d$ ferromagnet to spin glass
transition at $T\!=\!0$ can be explained in terms of a
percolation transition in the following way: For a spin glass, one starts with
a ferromagnetic configuration and searches for loops in the dual lattice
with negative weight. These loops correspond to clusters of spins, which
can be flipped to decrease the energy. If these loops are small, i.e.
not percolating, the ground state is ferromagnetic, otherwise
the ground state exhibits spin-glass order. 
A different argument relating this transition to a percolation
transition was also briefly mentioned in a study, which 
focuses on the critical slowing down of polynomial-time algorithms 
at $T\!=\!0$ phase transitions \cite{middleton2002}.

Right at $\rho_c$ we studied $2d$ (hex, $3d$) systems up to $L\!=\!512$ ($768$, $96$) 
with
$3.2\!\times\!10^{4}$ ($1.6\!\times\!10^{3}$, $6.4\!\times\!10^{3}$)  samples.
For the $2d$ systems we found $\gamma$ listed in table
\ref{tab:crit_exp} with  qualities $Q^{\pm}_{\rm 2d}\!=\!0.61$, 
$Q^{\pm}_{{\rm hex}}\!=\!0.10$ and
$Q^{\rm{Gl}}_{\rm 2d}\!=\!0.51$.  In $3d$, due to the very small percolation
probability $0.0014(1)$ at $\rho_c$ our  results are less clear. 
Hence, we can draw no final conclusions for that case.
Regarding the fractal dimension, spanning loops
and the non-spanning loops exhibit within error bars the same
exponents $d_f=\tilde{d}_f$ in $2d$ (values of $d_f$ listed in 
table \ref{tab:crit_exp}). As in the MWP case, 
the non-spanning loops are compact 
($\langle v\rangle\!\sim\!R^2$).
On the other hand, in $3d$, 
we find $\tilde{d}_f=1.43(2)$ different from $d_f=1.30(1)$.

Also regarding the mean weight $\langle \omega_\ell \rangle$
for given length $\ell$, we find in
$2d$ that spanning and non-spanning loops exhibit  both $\langle
\omega_\ell \rangle\!\sim\!\ell$ (i.e. 
$\sim L^{d_f}$ for the spanning loops), while in $3d$ the quality of
the data is again not sufficient to observe a clear power law.
%

\begin{figure*}[t]
\centerline{\includegraphics[width=0.9\linewidth]{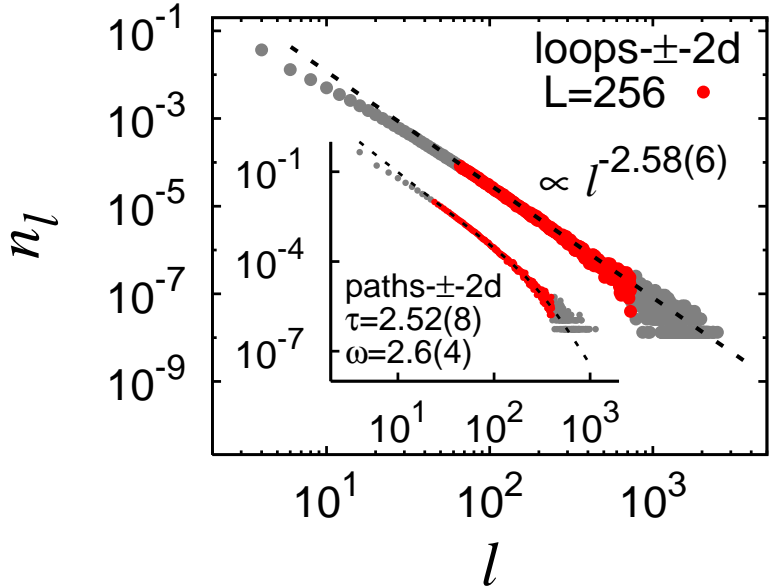}}
\caption{ Results for the $2d$ $\pm$J loops:
distribution $n_\ell$ of the loop-lengths $\ell$ at criticality,
excluding the spanning loops (main plot), gray data points were
omitted from the fit. The inset shows $n_\ell$ for the case of MWPs, 
where the fit accounts for corrections to scaling (see text).
\label{fig:len_distrib}}
\end{figure*}  

\begin{figure}[t]
\centerline{ \includegraphics[width=0.9\linewidth]{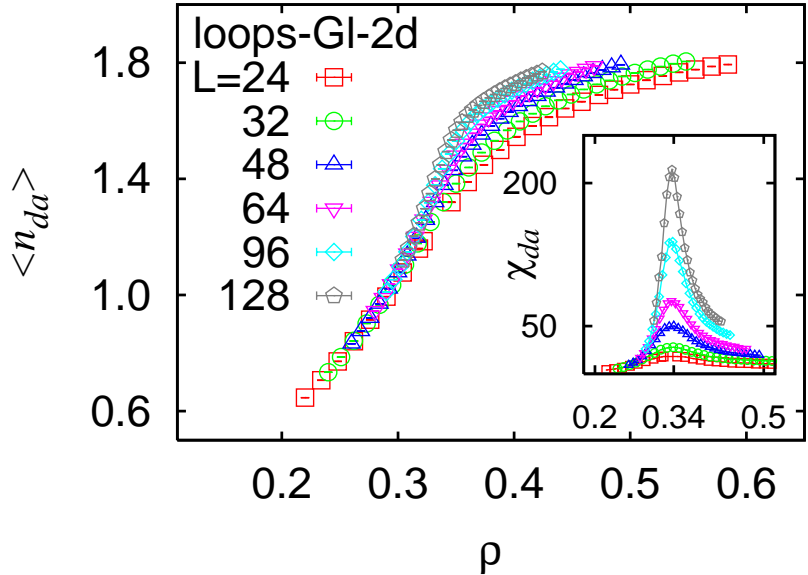} }
\caption{ Performance of the algorithm for $2d$ Gl loops: 
the main plot shows the average number of dual adjustments per
site for different system sizes $L$ and the inset illustrates 
the associated finite-size susceptibility.
\label{fig:performance}}
\end{figure}  

The distribution of loop-lengths, excluding the truly spanning
loops, is  in $2d$ in good agreement with a power-law decay
$n_\ell\!\sim\!\ell^{-\tau}$, see figure \ref{fig:len_distrib} and 
table \ref{tab:crit_exp}.  
In $3d$ we find again strong finite size effects and hence a most reliable 
estimate of $\tau$ using a scaling form 
$n_\ell\!\sim\!\ell^{-\tau}/(1+b\ell^{\omega})$ .

Finally, we address the performance of the algorithm.
We doe not want to measure the running time in terms
of CPU minutes, since this is machine dependent and is
also influenced by external
factors like which other processes are running. Hence, we
have to look at the algorithm more closely.
So as to determine a minimum-weight perfect matching,
the algorithm attempts to find an optimal solution to an associated
\emph{dual} problem \cite{cook1999,opt-phys2001}. Therein, the 
basic task of the algorithm is to find \emph{augmenting} paths that 
improve the solution to the latter. While executing, the solution to 
the dual problem is adjusted several times. As a measure for the 
algorithm performance we consider the number $N_{da}$ of such 
adjustment operations, until the algorithm terminates. 
Figure \ref{fig:performance} shows the average number of these 
operations per lattice site 
$\langle n_{da} \rangle\!=\!L^{-d}\langle N_{da} \rangle$ 
for the case of a Gaussian-like distribution of the edge weights.
The value of $\langle n_{da} \rangle$ increases with increasing $\rho$ and 
curves for different system sizes deviate from each other as the 
critical point 
is approached from below. At $\rho_c$ it scales like 
$\langle n_{da} \rangle \!\sim\!L^{2.074(4)}$.  
Moreover, the corresponding susceptibility $\chi_{da}$, i.e.\
the fluctuations of the $ n_{da}$, diverges at
the critical point  where is scales as $\chi_{da}\!\sim\!L^{1.41(4)}$, 
as can be seen from the inset of figure \ref{fig:performance}. 
For the case of a $\pm$J distribution of the edge weights, we
found similar but slightly different results (not depicted):
 Here, data sets that describe the scaling of 
$\langle n_{da} \rangle$ for different system sizes, fall onto 
one universal curve everywhere. 
This holds also for $\chi_{da}$, which is 
peaked slightly below the critical point at $\rho\!\approx\!0.096$.


\section{Conclusions \label{sect:conclusions}}
We have introduced a percolation problem, where edge weights with possibly
negative values are attached to the edges. A system is called percolating
if spanning loops or paths of negative weight exist. Hence, this
model might be fundamentally different from classical percolation problems.
We studied systems in $2d$ and $3d$ up to large system sizes. In all cases,
the universality class is indeed clearly different from standard bond/site
percolation. 
 Further, the values of the fractal 
exponents obtained for the  paths/loops differ from those reported in 
the context of shortest paths or hulls of percolation clusters, so far.
Nevertheless, the numerical values for the exponents found here are close
to those of disorder-induced single defects ($d_f\!=\!1.261(16)$) and 
multiple defects ($d_f\!=\!1.250(3)$) in a $2d$ elastic medium at zero
temperature \cite{middleton2000}.
We observe universality in $2d$, hence the properties are independent of 
the type of weight distribution,
lattice geometry
and the same for loops or paths. 
We also studied (not shown here) a related problem, where the loops 
are allowed to intersect.
The corresponding mapping was recently used in a different context
to determine extended ground states of Ising spin glasses \cite{thomas2007}.
We found again the same critical exponents.
On the other hand, we observe in $2d$ the same
behavior as for the $T\!=\!0$ ferromagnet to spin glass transition for the
random-bond Ising model, hence this physical transition can be probably
explained by a percolation transition in the spirit of the model
we have introduced here. Hence, studying percolation problems
with negative weights and similar generalizations 
might be a key approach to describe many yet not well-understood
phase transitions in terms of percolation transitions. We also
anticipate applications to other fields like social problems (see introduction)
or other types of networks.

\ack{
We acknowledge financial support from the VolkswagenStiftung (Germany)
within the program  ``Nachwuchsgruppen an Universit\"aten''. The
simulations were performed at the  ``Gesellschaft f\"ur
Wissenschaftliche Datenverarbeitung'' and 
the ``Institute for Theoretical Physics'', both in G\"ottingen
(Germany).}


\section*{References}

\begin{thebibliography}{10}

\bibitem{comment_cookrohe}
For the calculation of minimum-weighted perfect matchings we use Cook and Rohes   
blossom4 extension to the Concorde library
{\tt http://www2.isye.gatech.edu/\~{}wcook/blossom4/}

\bibitem{ahuja1993}
Ahuja R K, Magnanti T L and Orlin J B 1993
{\it {Network Flows: Theory, Algorithms and Applications}}
(New Jersey: {Prentice Hall})

\bibitem{amoruso2004}
Amoruso C and Hartmann A K 2004
{\it Phys. Rev.} B {\bf 70} 134425

\bibitem{antunes1998}
Antunes N D and Bettencourt L M A 1998
{\it Phys. Rev. Lett.} {\bf 81} 3083--86

\bibitem{bittner2005}
Bittner E, Krinner A and Janke W 2005
{\it Phys. Rev.} B {\bf 72} 094511

\bibitem{cook1999}
Cook W and Rohe A 1999
{\it INFORMS J. Computing}  {\bf 11} 138--48

\bibitem{engels1996}
Engels J, Mashkevich S, Scheideler T and Zinovjev G 1996
{\it Phys. Lett.} B {\bf 365} 219

\bibitem{opt-phys2001}
Hartmann A K and Rieger H 2001
{\it Optimization Algorithms in Physics}
(Berlin: Wiley-VCH)

\bibitem{s_value}
Houdayer J and Hartmann A K 2004
{\it Phys. Rev.} B {\bf 70} 014418
\newblock $S$ measures the mean square distance of the scaled data to the
  master curve in units of standart errors

\bibitem{melchert2007}
Melchert O and Hartmann A K 2007
{\it Phys. Rev.} B {\bf 76} 174411

\bibitem{pfeiffer2003}
Pfeiffer F O and Rieger H 2003
{\it Phys. Rev.} E {\bf 67} 056113 

\bibitem{schakel2001}
Schakel A M J 2001
{\it Phys. Rev.} E  {\bf 63} 026115

\bibitem{stauffer1994}
Stauffer D and Aharony A 1994
{\it Introduction to Percolation Theory}
(London: Taylor and Francis)

\bibitem{wenzel2005}
Wenzel S, Bittner E, Janke W, Schakel A M J and Schiller A 2005
{\it Phys. Rev. Lett.} {\bf 95} 051601

\bibitem{thomas2007}
Thomas C K and Middleton A A 2007 
{\it Matching Kasteleyn cities for spin glass ground states} 
arXiv:0706.2866

\bibitem{middleton2000}
Middleton A A 2000 
{\it Phys. Rev.} B {\bf 61} 14787

\bibitem{middleton2002}
Middleton A A 2002
{\it Phys. Rev. Lett.} {\bf 88} 017202

\end{thebibliography}
\end{document}